\PassOptionsToPackage{colorlinks=true,citecolor=blue,urlcolor=blue,linkcolor=blue}{hyperref}
\documentclass[aps,prl,reprint,amsmath,amssymb,superscriptaddress, 10pt]{revtex4-2}

\usepackage{amsmath,amssymb}
\usepackage{graphicx}
\usepackage{braket}
\usepackage[normalem]{ulem}
\usepackage{enumerate}
\usepackage{MnSymbol}
\usepackage{listings}
\usepackage[shortlabels]{enumitem}
\usepackage{orcidlink}

\newcommand{\be}{\begin{equation}}
\newcommand{\ee}{\end{equation}}

\newcommand{\para}[1]{\paragraph{\noindent\textbf{#1---}}}

\newcommand{\dd}{{\rm d}}
\newcommand{\dr}{^\text{dr}}

\newcommand{\eff}{\text{eff}}

\newcommand{\pr}{{\text{pth}}}
\newcommand{\thr}{{\text{th}}}
\newcommand{\pp}{{\chi}}

\begin{document}

\newcommand{\titleinfo}{Universal efficiency boost
in prethermal quantum heat engines at negative temperature}

\title{\titleinfo}

\author{Alberto Brollo\orcidlink{0009-0004-4939-516X}} 
\email{alberto.brollo@tum.de}
\affiliation{Technical University of Munich, CIT, Department of Mathematics, Boltzmannstraße 3, 85748 Garching, Germany}
\affiliation{Technical University of Munich, TUM School of Natural Sciences, Physics Department, 85748 Garching, Germany}

\author{Adolfo del Campo\orcidlink{0000-0003-2219-2851}}    
\email{adolfo.delcampo@uni.lu}
\affiliation{Department  of  Physics  and  Materials  Science,  University  of  Luxembourg,  L-1511  Luxembourg,  Luxembourg}
\affiliation{Donostia International Physics Center,  E-20018 San Sebasti\'an, Spain}

\author{Alvise Bastianello\orcidlink{0000-0003-3441-671X}} 
\email{alvise.bastianello@tum.de}
\affiliation{Technical University of Munich, TUM School of Natural Sciences, Physics Department, 85748 Garching, Germany}
\affiliation{Munich Center for Quantum Science and Technology (MCQST), Schellingstr. 4, 80799 M{\"u}nchen, Germany}

\begin{abstract}
Heat engines near the adiabatic limit typically assume a working medium at thermal equilibrium. However, quantum many-body systems often showcase conservation laws that hinder thermalization, leading to prethermalization in exotic stationary phases. This work explores whether prethermalization enhances or reduces engine efficiency. 
We investigate Otto cycles in quantum systems with varying numbers of conserved quantities. We find that additional conservation laws reduce efficiency at positive temperatures, but enhance it in regimes of negative temperatures. Our findings stem from general thermodynamic inequalities for infinitesimal cycles, and we provide evidence for integrable models undergoing finite cycles using the theoretical framework of Generalized Hydrodynamics. The relevance of our results for quantum simulators is also discussed.
\end{abstract}

\maketitle

The analysis of heat engines has played a key role since the birth of thermodynamics \cite{Callen1991}. The advent of quantum thermodynamics has followed a similar path, with the design and characterization of quantum heat engines \cite{Mahler2014,Alicki2018}. Early theoretical proposals \cite{Alicki1979} have been adapted for their implementation with current platforms for quantum technologies, including trapped ions \cite{Rossnagel2016,Maslennikov2019,Zhang2022}, nitrogen-vacancy centers \cite{Klatzow2019}, ultracold gases \cite{Koch2023}, and NMR systems \cite{Peterson2019}. 
Technological advances have motivated studies beyond canonical equilibrium involving coherence, squeezing, negative temperatures \cite{Scully03,Rossnagel2014,Niedenzu2018,deAssis2019,Nettersheim2022}, and genuine nonequilibrium protocols, although such processes typically reduce efficiency due to irreversibility.  
Driving schemes such as shortcuts to adiabaticity \cite{Campo2014,Beau2016,delCampo2018} fast-forward a quantum adiabatic evolution in finite time, but their exact implementation can be challenging \cite{Hartmann2020} and aims for the same efficiency of adiabatic thermal cycles.

Designing quantum heat engines utilizing many-body systems as a working medium is necessary for their scaling \cite{Jaramillo2016,Beau2016} and paves the way to harness a wide variety of phenomena without a single-particle counterpart, including quantum statistics \cite{Kim2011,Watanabe2020}, interparticle interactions \cite{Chen2019,Bouton2021,Koch2023,Watson25}, and critical phenomena \cite{Campisi2016}.
As many-body systems generally thermalize, the working medium follows equilibrium states if subject to slow operations. Hence, most previous studies have considered working medium at thermal equilibrium.
Yet, several many-body systems feature constraints that forbid canonical thermalization \cite{Ueda2020} and present genuine prethermal phases in which quantum simulators \cite{Schmiedmayer2018,Eisert2021} could perform reversible operations. A natural question is whether this scenario could be advantageous in increasing the performance of quantum heat engines. An important precedent in this regard is the study of quantum heat engines that harness many-body localization \cite{Halpern2019}.

\begin{figure}[b!]
\includegraphics[width=0.7\columnwidth]{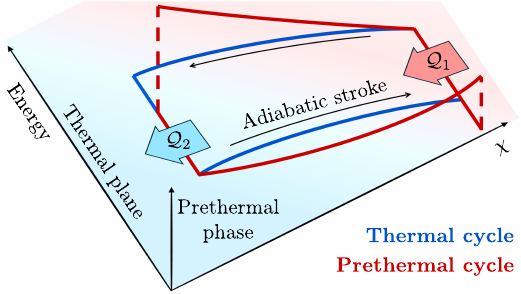}
\caption{
\textbf{Thermal vs prethermal Otto cycles.} 
The thermodynamic ensemble (thermal plane) describing the adiabatic strokes of a thermalizing working medium is described by a few conserved charges, including the energy, and the control parameter $\pp$. Prethermal working matter is characterized by a larger number of charges, exploring prethermal phases.
}\label{Fig_cartoon}
\end{figure}

This Letter investigates the impact of \emph{prethermalization} \cite{Ueda2020} in extended many-body systems on the engine efficiency, revealing that its advantage,  relative to thermalizing working media, is universally determined by the temperatures of the thermal baths. In particular, prethermal heat engines that operate at negative temperatures exhibit a universal efficiency boost. 

We consider Otto cycles operating between two thermal baths and performing work through a tunable parameter $\pp$. Without altering the baths, we compare prethermalization against thermalization during the adiabatic strokes, highlighting the role of the many-body working medium, whereas previous works focused on prethermal baths in few-body engines \cite{Manzano2020}.
We consider the following strokes depicted in Fig. \ref{Fig_cartoon}: 
\begin{enumerate}[(i),noitemsep,nolistsep,leftmargin=0.4cm]
\item \emph{Adiabatic transformation.} The system starts at thermal equilibrium with the bath, then it evolves in isolation, slowly changing $\pp$ and possibly exploring prethermal phases. It exchanges work $\mathcal{W}_1$, but not heat, and the entropy is conserved.
\item\emph{Isochore transformation.} At constant $\pp$, the system reaches thermal equilibrium with the second bath exchanging heat $\mathcal{Q}_1$, but not work.
\item The cycle is closed with another adiabatic stroke exploring a prethermal phase, exchanging work $\mathcal{W}_2$, and a isochoric transformation, exchanging heat $\mathcal{Q}_2$.
\end{enumerate}

The system is governed by a Hamiltonian of the form $H(\pp)=H_0(\pp)+\epsilon V$, where the perturbation governed by the small parameter $\epsilon$ governs prethermalization.
 Specifically, we assume
the unperturbed Hamiltonian $H_0(\pp)$ features $N$ conserved charges $\{Q_j\}_{j=1}^N$, which may depend on $\chi$, and satisfy $[Q_j, H_0(\pp)]=0$. 
The small perturbation $\epsilon V$ is assumed to break some charges and deforms the others $Q_j\to Q_j'=Q_j+\delta Q_j(\epsilon)$, in such a way that $[Q_j', H(\pp)]=0$ for $j\in\{1,...,N'<N\}$. 
This is a fairly common scenario. For example, resonant 
tunneling in multicomponent ultracold gases \cite{Lannig2020} or tunneling in adjacent superfluids \cite{Schweigler2017} breaks inter-species particle number conservation, Floquet-engineered Hamiltonians showcase symmetries broken at finite driving frequencies \cite{Bukov2015}, and integrable systems \cite{Calabrese2016,Guan2022} have infinitely many conservation laws broken by perturbations.

Isolated quantum systems with weakly-broken conserved charges undergo prethermalization: after a fast timescale $t_\pr$ decided by microscopic processes \cite{Calabrese2011}, the system relaxes to the maximally-entropic state compatible with the conservation laws of the unperturbed Hamiltonian $H_0$. This is the prethermal phase. Then, on a slow time scale $t_\thr\gg t_\pr$ usually polynomially growing in $\epsilon^{-1}$ \cite{Bertini2015,Mallayya2021,Surace2023}, relaxation to a steady state determined by the reduced set of charges is observed. This is the thermal phase, where only few charges like the Hamiltonian and the particle number are conserved.
For commuting charges, the two phases are described by generalized Gibbs ensemble (GGE) $\hat{\rho}= Z^{-1} e^{-\sum_j \beta_j Q_j}$ \cite{Rigol2007,Calabrese2016}\, where the appropriate charges and generalized inverse temperatures $\beta_j$ are considered. Although the GGE excludes certain nonergodic mechanisms, such as many-body localization \cite{Abanin2019,Halpern2019} and fractons \cite{Moudgalya2022}, it remains very general. 

To highlight the impact of conservation laws only, we consider adiabatic processes where the adiabatic strokes follow  the appropriate GGE, being it  prethermal $(t_\pr\ll \chi/(\dd\chi/\dd t)\ll t_\thr)$ or thermal $(t_\thr\ll \chi/(\dd\chi/\dd t))$, and compare the two cases in the small $\epsilon$ limit. This adiabaticity requirement is much less stringent than quantum adiabaticity \cite{Berry2009}, whose time scale diverges in the absence of an energy gap, as is common in the thermodynamic limit.
Our focus is on the cycle's efficiency, namely the ratio between the total work $\mathcal{W}=\mathcal{W}_1+\mathcal{W}_2$ and the absorbed heat $\mathcal{Q}_\text{abs}=\max(\mathcal{Q}_1,\mathcal{Q}_2)$
\be
\eta=\mathcal{W}/\mathcal{Q}_\text{abs}\, .
\ee 
We unveil a \emph{universal efficiency enhancement}: Thermalizing medium is more efficient for a positive bath temperature, whereas prethermalization is convenient at a negative temperature.
This holds whenever all the charges conserved by the thermalizing dynamics, with the exception of the Hamiltonian $H_0(\pp)$, are independent of the control parameter $\pp$, while the whole set of prethermal charges can depend on $\pp$.
We provide analytical proof for infinitesimal cycles on the basis of general thermodynamic inequalities without any assumption on the number of conservation laws, the form of interactions, or the dimensionality of the system. We furthermore demonstrate our findings using finite cycles with integrable systems, i.e., minimal interacting one-dimensional models featuring infinitely many conservation laws, amenable to many-body analytical computations far from equilibrium \cite{Calabrese2016,Alvaredo2016,Bertini2016,Bastianello2022}. 
An interaction-driven quantum Otto cycle has been experimentally realized in a three-dimensional atomic cloud across the BEC-BCS crossover \cite{Koch2023}, and nearly-integrable variants are also possible \cite{Chen2019}.
However, realizing negative temperatures requires Hamiltonians with a finite maximum energy: this is not possible in continuous systems, but it is conceivable in experiments on a lattice \cite{Medley2011,Braun2013,deAssis2019}. 
As a proof of concept, we concretely discuss how our findings can be probed in state-of-the-art quantum gas microscopes, realizing integrable spin chains \cite{Wei2022} with tunable integrability-breaking perturbations. In this context, we discuss how to engineer negative temperature states and measure the work done during the (pre)thermal adiabatic strokes.

\section{Results}

\para{The adiabatic flow equations.} 
We begin studying the evolution of the state during the adiabatic strokes. In the limit of slow changes of the control parameter $\chi$, the system instantaneously follows a GGE with evolving generalized temperatures, which we conveniently arrange in a vector $\{\beta_j\}_j\to \vec{\beta}$.
As advanced in the introduction, we consider the limit where the perturbation breaking conservation laws is infinitesimal $\epsilon\to 0$, such that its effect is solely to break certain charges, without substantially altering the remaining ones.

We derive the \emph{flow equations} that govern the adiabatic strokes in the prethermal states, since the thermal case follows similarly. 
To this end, it is convenient to approximate the smooth evolution as a sequence of sudden increments $\pp\to \pp+\dd\pp$ separated by a waiting time $\dd t$. The adiabatic limit $\dd \pp/\dd t\to 0$ is then taken, considering a large waiting time, in such a way that the system prethermalizes to the new generalized inverse temperatures $\vec{\beta}^\pr(\pp+\dd\pp)$.
Let $Q_j(\pp)$ be the parametrically $\pp-$dependent charge, and $\langle...\rangle_{\pp, \vec{\beta}}$ be the expectation value in the prethermal state at $\pp$. The parameters $\vec{\beta}^\pr$ are determined by charge conservation $\langle Q_j(\pp+\dd \pp)\rangle_{\pp, \vec{\beta}^\pr(\pp)}=\langle Q_j(\pp+\dd \pp)\rangle_{\pp+\dd \pp, \vec{\beta}^\pr(\pp+\dd\pp)}$. To linear order, one gets the flow equations
\be\label{eq_flow}
C_\pr\partial_\pp \vec{\beta}^\pr+A_\pr\vec{\beta}^\pr=0\, ,
\ee
where the $\pp-$dependence is omitted to ease the notation. Above, we defined the \emph{static covariance matrix} as the connected charge-charge correlators $[C_\pr]_{i,j}=\langle Q_i Q_j\rangle_\text{c}$, and the \emph{susceptibility matrix} $[A_\pr]_{i,j}=\langle Q_i \partial_\pp Q_j\rangle_\text{c}$. Here we defined the connected correlators as $\langle\mathcal{O}_1\mathcal{O}_2\rangle_\text{c}\equiv\langle\mathcal{O}_1\mathcal{O}_2\rangle-\langle\mathcal{O}_1\rangle\langle\mathcal{O}_2\rangle$. 
In addition, Eq. \eqref{eq_flow} implies the adiabatic evolution of the charges 
\be\label{eq_ch_flow}
\partial_\chi \langle Q_j\rangle=\langle \partial_\chi Q_j\rangle\,. 
\ee

The derivation of Eqs. \eqref{eq_flow} and \eqref{eq_ch_flow} is reported in Methods.
The thermal flow equations are identical to Eq. \eqref{eq_flow}, restricted to the proper conserved charges and inverse temperatures $\vec{\beta}^\thr$. 

To relate the flow equations \eqref{eq_flow} and the conventional notion of adiabaticity based on entropy, it is instructive to define the free energy associated with the GGE through the partition function as $\mathcal{F}=-\log\text{Tr}\left[ e^{-\sum_j\beta_j Q_j}\right]$, from which the entropy $S$ is defined through canonical thermodynamic identities $\mathcal{F}=\sum_j \beta_j \langle Q_j\rangle-S$. By taking the time derivative of the last expression and comparing it with $\tfrac{\dd \mathcal{F}}{\dd t}$ obtained from the partition function, one reaches
\be
\frac{\dd S}{\dd t}=\frac{\dd \chi}{\dd t}\sum_j \beta_j\big(\partial_\chi \langle Q_j\rangle-\langle \partial_\chi Q_j\rangle\big)\, .
\ee

Notice that Eq. \eqref{eq_ch_flow} \emph{implies} $\tfrac{\dd S}{\dd t}=0$, and \emph{it follows} from the flow equations \eqref{eq_flow} that the adiabatic strokes are reversible in the conventional sense, regardless of whether the working medium is described by a thermal or prethermal phase. 
Therefore, both adiabatic strokes done with a thermalizing and prethermalizing medium belong to the class of reversible operations: our goal is now to understand which choice benefits the engine's efficiency.
Eq. \eqref{eq_flow} is difficult to solve as it is highly nonlinear, since the expectation values evolve with the complex many-body state. Further progress can be made in generic infinitesimal cycles and in integrable models where the static covariance matrix $A$ and the susceptibility matrix $C$ can be analytically computed.
\para{Universal efficiency boost in infinitesimal cycles.} 
Although infinitesimal cycles cannot be reliably used to deduce the behavior of finite cycles, they already provide a good indication. Furthermore, we found infinitesimal cycles to be amenable of \emph{analytical universal results}. Our findings are universal in the sense that they rely solely on general thermodynamic inequalities, without making any assumptions about the number of conserved quantities, the form of interactions, or the dimensionality of the system. Hence, they have the broadest applicability.

We consider two thermal baths at $\vec{\beta}$ and $\vec{\beta}+\delta \vec{\beta}$, and the two strokes running from $\pp$ to $\pp+\delta \pp$.
\begin{figure}[t!]
\includegraphics[width=0.99\columnwidth]{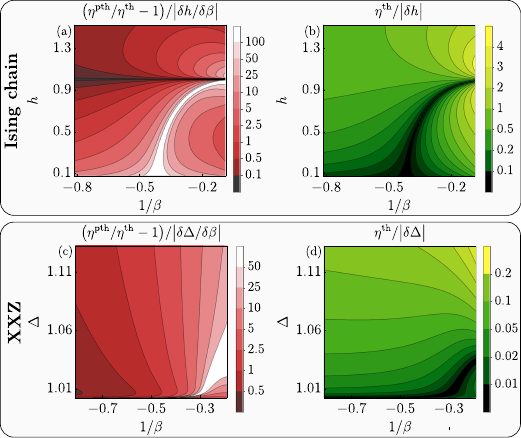}
\caption{\textbf{Thermal vs Prethermal infinitesimal Otto cycles in integrable models. } We show the relative efficiency of thermal-prethermal infinitesimal Otto cycles in the Ising \eqref{eq_Ising} and antiferromagnetic $J=-1$ XXZ \eqref{eq_XXZ} chains with average magnetization $\langle \sigma^z\rangle=0.45$. We use as tunable parameters the magnetization $h$ and anisotropy $\Delta$, respectively. 
To plot efficiencies independent of the cycle size, we focus on \emph{skewed} cycles where the absorbed heat is much larger than the work, $\mathcal{W} \ll \mathcal{Q}_\text{abs}$, or equivalently $|\delta\chi|\ll|\delta\beta|$. In this regime, the efficiency scales as  $\eta\sim \delta \chi$ and $\eta^\pr/\eta^\thr - 1\sim \delta \chi/\delta \beta$, and the data is rescaled to factor out the explicit dependence on the size of the infinitesimal stroke. Explicit formulas are obtained from Eq. \eqref{eq_Den} and reported in SI \cite{suppmat}. 
Notice that the regions of large relative efficiency are related to regions of vanishing thermal efficiency.}
\label{Fig_inf_cycles}
\end{figure}
The change in internal energy during a stroke is obtained by expanding the integrated equation \eqref{eq_ch_flow}, where we consider the Hamiltonian as the conserved charge $\Delta \langle H\rangle=\int_{\pp}^{\pp+\delta \pp}\dd \pp' \langle \partial_{\pp'}H\rangle_{(\pp',\vec{\beta}(\pp'))}$. We find
\be\label{eq_Den}
\Delta \langle H\rangle=
\delta \pp \langle\partial_{\pp}H\rangle+\tfrac{(\delta\pp)^2}{2}\left[\partial_\pp\langle \partial_\pp H \rangle+\partial_\pp \beta_j\partial_{\beta_j}\langle \partial_\pp H\rangle\right]\, ,
\ee
\begin{figure*}[t!]
\includegraphics[width=0.99\textwidth]{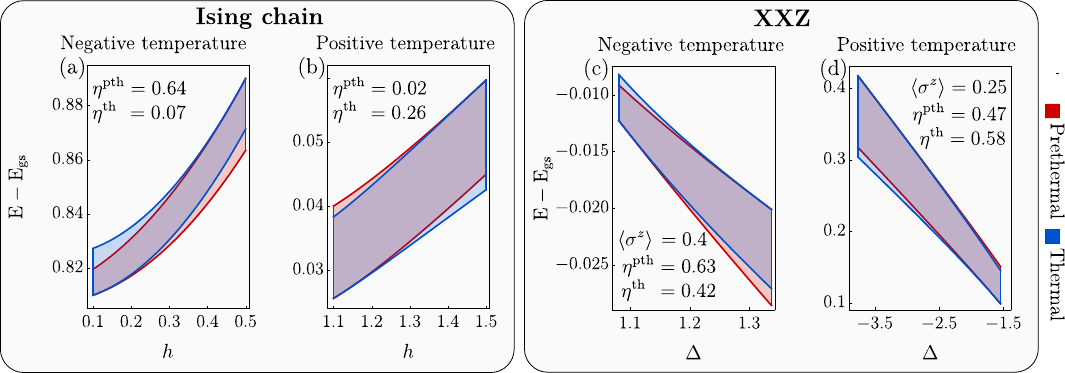}
\caption{\textbf{Finite Otto cycles in integrable models. }  
On the horizontal and vertical axes, we show the control parameter and the energy difference from the ground state (normalized with respect to the system size), respectively. In each panel, both cycles are represented and the values of the efficiencies are reported. Panels (a,b): Ising chain with cold(hot) reservoirs at temperatures $\beta^{-1}_\text{C}$($\beta^{-1}_\text{H}$) respectively.
Specifically, in (a) we consider negative temperatures
$(\beta^{-1}_\text{C},\beta^{-1}_\text{H})=(-0.70,-0.69)$ and in (b) positive temperature $(\beta^{-1}_\text{C},\beta^{-1}_\text{H})=(0.30,0.48)$. Panels (c,d): analog cases for the XXZ chain \eqref{eq_XXZ} with $J=-1$ and fixed magnetization $\langle\sigma^z \rangle$. The temperatures of the baths in (c) and (d) are $(\beta^{-1}_\text{C},\beta^{-1}_\text{H})=(-0.175,-0.150)$ and $(\beta^{-1}_\text{C},\beta^{-1}_\text{H})=(0.5,2.0)$.
These examples show that the general conclusions for infinitesimal cycles remain valid also for finite operations. A quantitative measure of the GGE's departure from canonical equilibrium and further cases are reported in SI \cite{suppmat}.
}\label{Fig_finite_cycles}
\end{figure*}
where repeated indices are summed and terms $\mathcal{O}(\delta \pp)^3$ are neglected. Expectation values are computed on the initial thermal state, and the choice of flow equations \eqref{eq_flow} determines the evolution through thermal or prethermal states.
Work is obtained by summing the two contributions from $\vec{\beta}$ and $\vec{\beta}+\delta \vec{\beta}$, and expanding in $\delta \vec{\beta}$.
Notice that the flow equation \eqref{eq_flow} does not affect the first order $\sim \delta \pp$ contribution, but only the second order $\sim \delta\pp^2$. Therefore, differences in the two cycles appear at order $\sim\delta \pp^2$.
Combining Eq. \eqref{eq_flow} and the identity $\partial_{\beta_j}\langle \partial_\pp H\rangle=-\langle Q_j \partial_\pp H\rangle_{\text{c}}$, the work difference $\delta \mathcal{W}\equiv \mathcal{W}^\pr-\mathcal{W}^\thr$ is determined by the covariance and susceptibility matrices. The result is further simplified if all charges in the thermal state, except for the Hamiltonian itself, are $\pp-$independent. In fact, $A_{i,j}$ vanishes for all indices $j$ of the charges of the thermalizing dynamics, with the exception of the Hamiltonian itself. $A_{i,j}$ may be non-zero for other prethermal charges, but these do not couple to the initial bath $\vec{\beta}$ and lead to
\be\label{eq_DW}
\delta \mathcal{W}=-\beta (\delta\pp)^2 \left[A_\pr^T C_\pr^{-1}A_\pr-A_\thr^T C_\thr^{-1}A_\thr\right]_{H,H}\, ,
\ee
where  $[...]_{H,H}$ denotes the diagonal element in the Hamiltonian direction, and $\beta$ is \emph{the canonical inverse temperature}.
The sign of Eq. \eqref{eq_DW} crucially depends on $\beta$, with $\delta\mathcal{W}$ having a sign opposite to $\beta$. In fact, the matrix product in Eq. \eqref{eq_DW} is reformulated within \emph{hydrodynamic projections} \cite{spohn2012large,Doyon2022} (see Methods) as the norm of a vector subtracting its projection onto a smaller subspace, which is always positive. 
After having considered the work difference, we now focus on the absorbed heat and note that both the thermal and prethermal strokes start with the same energy $\langle H\rangle_{\chi,\vec{\beta}}$. After the first adiabatic stroke, they will have reached internal energies $\langle H\rangle_{\chi,\vec{\beta}}+\mathcal{W}_1^{\thr\text{, }\pr}$. The absorbed heat is then computed as the difference in internal energy determined by the second bath, identical for the two working media, and the end of the adiabatic protocol $\mathcal{Q}_\text{abs}^{\thr\text{, }\pr}=\langle H\rangle_{\chi+\delta \chi,\vec{\beta}+\delta\vec{\beta}}-\langle H\rangle_{\chi,\vec{\beta}}-\mathcal{W}_1^{\thr\text{, }\pr}$. Inspection of Eq. \eqref{eq_Den} shows that $\mathcal{W}_1^{\pr}-\mathcal{W}_1^{\thr}=\mathcal{W}_2^{\pr}-\mathcal{W}_2^{\thr}=\frac{1}{2}\delta \mathcal{W}$, whence it follows that $\mathcal{Q}_\text{abs}^\thr-\mathcal{Q}_\text{abs}^\pr=\frac{1}{2}\delta \mathcal{W}$.  
Therefore, the relative efficiency can be written as $\tfrac{\eta^\pr}{\eta^\thr}=\left(1+ \tfrac{\delta\mathcal{W}}{\mathcal{W}^\thr}\right)\left(1-\tfrac{\delta\mathcal{W}}{2\mathcal{Q}^\thr_\text{abs}}\right)^{-1}$ and it is thus greater than one for $\delta \mathcal{W}>0$, which implies $\eta^\pr>\eta^\thr$.  
We conclude that infinitesimal Otto cycles operating with thermal medium are more efficient than prethermal ones at positive temperatures, whereas the opposite holds for negative $\beta$. 
\para{Prethermal finite cycles in integrable models.} We next explore the persistence of the efficiency inequality for finite cycles in nearly integrable models \cite{Bertini2015,Bastianello2021}. These systems are realized in cold atoms \cite{Guan2022}, have infinitely many conserved charges, are strongly interacting, and are yet amenable to exact analytical computations, making them ideal candidates for our scope. In GGEs, the expectation values of charges, the covariance and susceptibility matrices can be computed exactly within the framework of Thermodynamic Bethe Ansatz (TBA) \cite{takahashi2005thermodynamics}; see Methods.
We focus on two prototypical examples: the Ising model
\be\label{eq_Ising}
H_\text{Ising}=-\sum_j \left(\sigma^x_{j+1}\sigma_j^x+h\sigma_j^z\right)\, ,
\ee
and the XXZ spin chain
\be\label{eq_XXZ}
H_\text{XXZ}=-J\sum_j \left(\sigma^x_{j+1}\sigma_j^x+\sigma_{j+1}^y\sigma_j^y+\Delta \sigma^z_{j+1}\sigma^z_j\right)\, ,
\ee
where $\sigma_j^{x,y,z}$ are canonical Pauli matrices acting on the $j$-th site. Notice that for the sake of simplicity, we work in dimensionless energy units. We consider homogeneous infinite systems and use as control parameters the magnetic field $h\to\pp$ and the anisotropy $\Delta\to\pp$ respectively. The Ising model is a canonical example of an integrable system, equivalent to noninteracting fermions with momentum $\lambda$ and dispersion $e(\lambda)=2\sqrt{(\cos \lambda-h)^2+\sin^2\lambda}$; see Supplementary Information (SI) \cite{suppmat} for details. We choose it primarily for pedagogical reasons \cite{Mbeng2024} and for its broad relevance from non-equilibrium physics \cite{Calabrese2011,Essler2016} to quantum engines \cite{Revathy20,Piccitto2022,Arezzo2024}. The XXZ chain is a paradigmatic example of an \emph{interacting} integrable model \cite{Franchini2017} implemented in quantum simulators \cite{Jepsen2020,Wei2022,Rosenberg2024,Hild2014,Scheie2021,Jepsen2022}; see SI \cite{suppmat} for details of its thermodynamics.
In the Ising chain, many-body eigenstates can be described as a gas of free fermionic quasiparticles with momentum density $\rho(\lambda)$ and extensive energy $\langle H_\text{Ising}\rangle=L\int \dd\lambda\,  e(\lambda)\rho(\lambda)$, with $L$ the system size \cite{Mbeng2024,suppmat}, where we removed the ground state energy. Choosing the function $\rho(\lambda)$ is equivalent to fix the parameters $\beta_j$ of the GGE \cite{Calabrese2016}, see Methods and SI \cite{suppmat}. On thermal states, it has Fermi-Dirac statistics $\rho(\lambda)=\tfrac{1}{2\pi}(e^{\beta e(\lambda)}+1)^{-1}$. 
The same picture holds in the XXZ chain, albeit interactions dress the excitations and deform thermal distributions through nonlinear integral equations \cite{suppmat}.
The flow equations Eq. \eqref{eq_flow} can be reformulated in the quasiparticle basis within Generalized Hydrodynamics (GHD) \cite{Alvaredo2016,Bertini2016,Bastianello2022,Doyon2025}, a non-perturbative framework for nearly integrable systems. In homogeneous systems with slow-varying interactions, the GHD equations are \cite{Bastianello2019}
\be\label{eq_ghd}
\partial_\pp \rho(\lambda)+\partial_\lambda(F^\text{eff}(\lambda)\rho(\lambda))=0\, .
\ee
The effective force $F^\text{eff}$ captures the many-body effect of varying interactions, and it vanishes for noninteracting models such as the Ising chain; see Methods for details.
In Fig. \ref{Fig_inf_cycles}, we show the efficiencies for infinitesimal cycles for Ising and XXZ chains, in a broad spectrum of parameters. Thermal states in the Ising chain are determined solely by the energy. By contrast, in the XXZ model we consider integrability-breaking perturbations that preserve the total magnetization $\sum_j \sigma_j^z$, resulting in thermal states with two conserved quantities. Notice that the total magnetization does not depend on the anisotropy, i.e., our control parameter. As expected from our general argument, prethermal states are more efficient than thermal ones at negative temperatures. 
In Fig. \ref{Fig_finite_cycles}, we consider exemplificative cases of finite cycles numerically solving Eq. \eqref{eq_ghd} \cite{suppmat}. For finite strokes, the gain becomes comparable with the efficiency itself, and is generally larger for the Ising chain. This is due to the fact that thermal states in the XXZ chain have two conserved charges rather than one as in the Ising case, allowing them to be closer to the GGE. More details are provided in SI \cite{suppmat}.

\para{Quantum engine simulators.}  The perfect isolation of quantum simulators that makes long-time coherent dynamics and negative temperature possible hampers quantum engines requiring heat exchanges with thermal baths \cite{Schmiedmayer2018,Eisert2021,Chen2019,Nautiyal2024}. Nonetheless, current platforms can already probe the two adiabatic strokes of the Otto cycle separately.
The XXZ chain \eqref{eq_XXZ} with positive spin exchange $J>0$ is realized in one-dimensional gases at unit filling \cite{Duan2003}, encoding the $z-$direction of the spin in two hyperfine levels. Convenient platforms for our scopes are Lithium-based implementations in optical lattices \cite{Jepsen2020}, where $\Delta$ is tunable thanks to a Feschbach resonance, and quantum gas microscopes \cite{Wei2022} due to their ability to conduct single-site measurements and operations. The current XXZ chain quantum microscope with Rubidium atoms \cite{Wei2022} lacks a Feshbach resonance, fixing $\Delta\simeq 1$. However, quantum microscopes with Lithium are also available \cite{Boll2016}, and could combine the advantages of the two platforms in the near future. 
The tunable transverse confinement efficiently breaks integrability, interpolating between a one-dimensional and a ladder geometry \cite{Wei2022}.
The use of thermal baths at negative temperature generally requires engineering for their preparation. In this regard, negative temperatures can be realized by selectively exciting spins in high-energy configurations and evolving them in the presence of integrability-breaking perturbations that induce thermalization. For example, such states for the XXZ chain in the ferromagnetic phase could be an antiferromagnetic spin arrangement.
Atom imaging provides snapshots of the $z-$magnetization, from which arbitrary $zz$ correlations can be obtained \cite{Wei2022}. Directly probing the energy requires measurements in the other spin directions as well, but adiabatic operations conveniently give direct access to energy differences through integration of Eq. \eqref{eq_ch_flow} which, for the case of a tunable $\Delta$, requires measuring $\langle\sigma_{j+1}^z\sigma_j^z\rangle$ only.
In the absence of a tunable $\Delta$, time-dependent magnetic traps $ H_\text{XXZ}+\sum_j B_j(t)\sigma^z_j$ can be used to exert work, as suggested in Ref. \cite{Nautiyal2024,Jaramillo2016,Eisert2021}.
Indeed, smooth traps break integrability weakly, resulting in long-lived prethermal states \cite{Bastianello2021,Caux2019,Bastianello2020}. This possibility, however, pivots (pre)thermalization timescales to the trap's size: on the typical sizes of a few tens of spins, a conservative estimation suggests timescales of various tens of spin-exchange times, challenging the present coherence time.
Instead, (pre)thermalization after homogeneous quenches in $\Delta$ requires $\sim5$ spin-exchanges \cite{Pozsgay2014}, and thus is more convenient.

\section{Discussion}
In this Letter, we unveiled the universal impact of conservation laws and prethermalization on quantum engines. 
By focusing on Otto cycles, we have established general thermodynamic inequalities showing how the relative efficiency of small cycles with thermal or prethermal working medium is entirely determined by the baths' temperature. Specifically, a thermal working medium is more efficient at positive temperatures, whereas prethermal media enhance the engine efficiency at negative temperatures. Although the use of negative temperatures generally requires population inversion, one can anticipate scenarios in which the benefits of their use outperform their cost, if not fundamentally, in practice.
We focused on integrable models as a concrete case of study, showing the persistence of our conclusions beyond small cycles, where Generalized Hydrodynamics provides exact quantitative insight into far-from-equilibrium quantum matter. Our findings are of direct relevance to state-of-the-art quantum simulators. 
While we thoroughly discussed the realization of the XXZ model in quantum gas microscopes, other platforms such as Rydberg atoms in optical tweezers \cite{Scholl2023}, superconducting qubits \cite{Morvan2022}, and trapped ions \cite{Kranzl2023} could also be employed. Fermi-Hubbard quantum microscopes with Feschbach resonances, offer another natural platform \cite{Boll2016} in the context of nearly-integrable models.
However, as shown by our analysis of infinitesimal cycles, our results are of broad relevance beyond integrability itself and apply generally whenever a conservation quantity can be selectively broken and are thus of broad experimental relevance. 
It is worth emphasizing that, albeit we focused on quasi-static protocols, (pre)thermalization time scales are driven by microscopic processed \cite{Calabrese2011,Bertini2015,Mallayya2021,Surace2023} that are usually fast compared with other typical scales in experiments \cite{Guan2022}, suggesting our results can also be predictive for finite-time protocols.
Interesting future directions involve exploring the consequences of nonthermal baths, which may be realized by coupling different portions of isolated quantum many-body systems \cite{Eisert2021}, and considering driving protocols in finite time, and their role on the tradeoff between the  efficiency and power, and (pre)thermalization.

\para{Data and code availability.}Raw data and working codes are available on Zenodo \cite{Zenodo}. 
\para{Acknowledgments.}
We are indebited to Manuele Landini, Rosario Fazio, Alvaro Martin Alhambra and Herbert Spohn for useful discussion and comments on the manuscript. 
 ABro acknowledges support from Deutsche Forschungsgemeinschaft (DFG, German Research Foundation) – TRR 352 – Project-ID 470903074. AdC acknowledges financial support from the Luxembourg National Research Fund (FNR Grant Nos. C24/MS/18940482/STAOpen). ABas acknowledges the support of the Deutsche Forschungsgemeinschaft (DFG, German Research Foundation) under the Germany's Excellence Strategy-EXC-2101-3990814868.
\para{Author contributions.} ABro carried out the analytical computations and simulations, ABas devised the project's goal. ABas and AC supervised the work. All authors contributed critically to the writing of the manuscript and the interpretation of results.

\section{Methods}
\para{Derivation of the flow equations.}
We derive the flow equations \eqref{eq_flow}, which govern the adiabatic evolution.
From the partition function $Z(\pp,\vec{\beta})=\text{Tr}\left[e^{-\sum_i \beta_i Q_i(\pp)}\right]$, one has the standard thermodynamic equalities $\langle Q_j(\pp)\rangle_{\pp,\beta_i}=-\partial_{\beta_i}\log Z(\pp,\vec{\beta})$, $\langle \partial_\pp Q_j(\pp)\rangle_{\pp,\beta_i}=-\partial_{\pp}\log Z(\pp,\vec{\beta})$, while the second mixed derivatives give connected correlation functions. Promoting $\vec{\beta}$ to be $\pp-$dependent in the adiabatic stroke, and imposing charge conservation $\langle Q_j(\pp+\delta \pp)\rangle_{\pp, \vec{\beta}(\pp)}=\langle Q_j(\pp+\delta \pp)\rangle_{\pp+\delta \pp, \vec{\beta}(\pp+\delta\pp)}$ to first order in $\delta\pp$, the flow equations Eq. \eqref{eq_flow} immediately follow.
Similarly, the generic variation of a charge can also be obtained with changes in $\pp$ and $\vec{\beta}_i$, leading to $\delta\langle Q_j \rangle =  \delta\pp\langle \partial_\pp Q_j\rangle
-\sum_i C_{ji} \delta\beta_i-\delta\pp\sum_i A_{ji} \beta_i$.
Imposing that $\vec{\beta}$ evolves with the flow equations, $\partial_\pp\langle Q_j \rangle=\langle \partial_\pp Q_j \rangle$ follows. 
\\

\para{Hydrodynamic projections.} Historically, hydrodynamic projections \cite{spohn2012large,Doyon2022} have been introduced to isolate the slow, long-wavelength dynamics of a many-body system by projecting onto conserved quantities. We use this framework to conveniently rewrite Eq. \eqref{eq_DW} and make its sign explicit. Due to the infinitesimal nature of the perturbation, the leading-order effect is the suppression of certain conserved quantities, rather than their deformation. Consequently, the set of conserved quantities under thermalizing dynamics spans a strict subspace of those preserved in the prethermal regime.
One introduces a scalar product in the vector space of the observable through their connected correlator $\langle \mathcal{O}_1|\mathcal{O}_2\rangle\equiv \langle \mathcal{O}_1\mathcal{O}_2\rangle_\text{c}$. We define the projector on the conserved charges of the (pre)thermal dynamics as $\mathbb{P^{\pr(\thr)}}=\sum_{i,j}^{N(N')}[C^{-1}_{\pr(\thr)}]_{i,j}|Q_i\rangle \langle Q_j|$, where the inverse static covariance matrix $C^{-1}$ is introduced for a properly normalized projection $\mathbb{P^{\pr(\thr)}}|Q_j\rangle=|Q_j\rangle$.
In this language, Eq. \eqref{eq_DW} is rewritten as the difference of the norm of a vector projected on different subspaces 
\be\label{eq_hydpr}
\delta\mathcal{W}= -\beta(\delta \chi)^2\left[\langle \partial_\pp H|\mathbb{P}^\pr|\partial_\pp H\rangle - \langle \partial_\pp H|\mathbb{P}^\thr|\partial_\pp H\rangle\right]
\,.
\ee
Since the space of thermal conserved charges is included in the prethermal one $[\langle \partial_\pp H|\mathbb{P}^\pr|\partial_\pp H\rangle - \langle \partial_\pp H|\mathbb{P}^\thr|\partial_\pp H\rangle\ge 0$, proving the sign of $\delta\mathcal{W}$ depends only on $\beta$.
\\

\para{Integrable systems.}
Multiparticle scattering events in integrable models are factorized into two-by-two elastic scattering events, entirely parametrized by their scattering phase.
In interacting integrable models, the \emph{rapidity} $\lambda$ and \emph{root density} $\rho(\lambda)$ generalize the momentum and momentum density of the free systems, respectively. An additional degree of freedom labeling quasiparticles of different species is present in several cases, such as in the XXZ model \cite{takahashi2005thermodynamics}, and is here omitted for brevity.
The expectation value of the conserved charges takes the form 
$\tfrac{1}{L}\langle Q_i \rangle = \int \dd\lambda \, q_i(\lambda) \rho(\lambda)$, with $q_i(\lambda)$ being the charge eigenvalue. The scattering phase $\Theta(\lambda-\lambda')$ between two excitations of rapidity $\lambda$ and $\lambda'$ depends on the rapidity difference. 
Thermal states and GGEs are described within the thermodynamic Bethe ansatz (TBA) framework \cite{takahashi2005thermodynamics}. More precisely, the state is parametrized by nonlinear integral equations
\be\label{eq_TBA}
\varepsilon(\lambda) = \sum_i \beta_i q_i(\lambda) + \int\frac{\dd\lambda'}{2\pi}\varphi(\lambda-\lambda')\log(1+e^{-\varepsilon(\lambda')})\, ,
\ee
where the scattering kernel is defined as $\varphi(\lambda)\equiv \partial_\lambda \Theta(\lambda)$. The pseudoenergy $\varepsilon(\lambda)$ parameterizes the state through the filling function $\vartheta(\lambda)=(1+e^{\varepsilon(\lambda)})^{-1}$, which is then connected to the root density as $\rho(\lambda)=\vartheta(\lambda)\frac{(\partial_\lambda p)\dr}{2\pi}$, with $p(\lambda)$ the momentum of a quasiparticle. In general, the dressing of a bare quantity $\tau(\lambda)$ is given as a solution of the linear integral equation $\tau\dr(\lambda) = \tau(\lambda) - \left[\varphi*\vartheta\tau\dr\right](\lambda)$. For brevity, we define the convolution $[\varphi*\tau](\lambda)=\int \dd\lambda'\varphi(\lambda-
\lambda')\tau(
\lambda')$. The static covariance matrix is analytically determined as
$\langle Q_iQ_j\rangle_\text{c}=-\partial_{\beta_i}\langle Q_j\rangle$ \cite{suppmat} 
\be\label{eq_chchcorr}
\tfrac{1}{L}\langle Q_iQ_j\rangle_\text{c}\overset{L\to\infty}{\simeq}\int \dd\lambda\, q_i\dr \rho(1-\vartheta)q_j\dr,
\ee
where $\simeq$ denotes equality in the thermodynamic limit.
The susceptibility matrix follows from $\langle Q_i\partial_\pp Q_j\rangle_\text{c}=-\partial_{\beta_i}\langle \partial_\pp Q_j\rangle$, where $\langle \partial_\pp Q_j\rangle$ is computed by means of the Hellmann-Feynman theorem \cite{Bastianello2019} 
\be\label{eq_HellF}
\tfrac{1}{L}\langle \partial_\pp Q_j\rangle\simeq\int \dd\lambda \left(\partial_\pp q_j \rho +\frac{1}{2\pi}\partial_\lambda q_jf\dr\vartheta\right),
\ee
where $f(\lambda) = -\partial_\pp p(\lambda) + \left[\partial_\pp\Theta*\vartheta(\partial_{\lambda} p)\dr\right](\lambda)$. Deriving Eq. \eqref{eq_HellF}, the susceptibility matrix follows \cite{suppmat} 
\be\label{eq_suscTBA}
\tfrac{1}{L}\langle Q_i\partial_\pp Q_j\rangle_\text{c}\simeq\int \dd\lambda\, q_i\dr \rho(1-\vartheta)\left(f\dr\frac{(\partial_\lambda q_j)\dr}{(\partial_\lambda p)\dr}-\Lambda_j\dr\right)
\ee
where $\Lambda_i(\lambda) = -\partial_\pp q_i(\lambda) + \left[\partial_\pp\Theta*\vartheta(\partial_{\lambda}q_i)\dr\right](\lambda)$. 
With the covariance and susceptibility matrices at hand, the flow equations \eqref{eq_flow} are fully determined. In the prethermal case, rather than working with infinitely many charges, it is more convenient to move to a quasiparticle basis. Here, the flow equations are equivalent to the GHD equations \eqref{eq_ghd} \cite{Bastianello2019} with the effective force being $F^\eff(\lambda)=f\dr(\lambda)/(\partial_\lambda p)\dr$, which can also be generalized to inhomogeneous setups. 
Notice that in non-interacting systems like the Ising model $\varphi=0$, therefore, the equations greatly simplify. In SI \cite{suppmat}, we provide details for the general formulas for the XXZ spin chain in the easy-axis regime $|\Delta|>1$. For $|\Delta|<1$, the GHD equations for changing $\Delta$ are an open problem \cite{Bastianello2019} that we do not address.

\para{Numerical methods.} Finite cycles in integrable systems are obtained by numerically solving the TBA and GHD equations; see SI \cite{suppmat} for details.  
The integral equations are discretized and solved with standard methods.
The GHD equation is solved using the method of characteristics at second order in time evolution \cite{Bastianello2019}.
The evolution along the thermal strokes is performed with the flow equations \eqref{eq_flow}.
We checked the convergence of our results with respect to the discretization in the rapidity space, the number of quasiparticle species in the XXZ chain, and the integration time step. Raw data and a Mathematica code are available on Zenodo \cite{Zenodo}.

\bibliography{biblio.bib}

\clearpage
\onecolumngrid
\newpage

\setcounter{equation}{0}  
\setcounter{figure}{0}
\setcounter{page}{1}
\setcounter{section}{0}    
\renewcommand\thesection{\arabic{section}}    
\renewcommand\thesubsection{\arabic{subsection}}    
\renewcommand{\thetable}{S\arabic{table}}
\renewcommand{\theequation}{S\arabic{equation}}
\renewcommand{\thefigure}{S\arabic{figure}}
\setcounter{secnumdepth}{2}  

\begin{center}
{\Large Supplementary Information\\
\titleinfo
} \\ 
Alberto Brollo, Adolfo del Campo, Alvise Bastianello
\end{center}

\bigskip

The Supplementary Information gathers the more technical, albeit standard, aspects of our Letter. In particular:

\begin{itemize}
\item Section \ref{sec_ising} gives a pedagogical overview of the Ising chain in the transverse field and provides further details on the infinitesimal and finite cycles presented in the main text.
\item Section \ref{sec_XXZ} reviews the thermodynamic Bethe ansatz approach to the XXZ model and further analyzes infinitesimal and finite cycles.
\item Section \ref{sec_num} discusses the numerical schemes used to solve the thermodynamics and hydrodynamics of integrable models.
\end{itemize}

\section{The quantum Ising chain in transverse field}
\label{sec_ising}
The one-dimensional transverse Ising model can be mapped onto a system of free spinless fermions via a Jordan-Wigner transformation. See Ref. \cite{Mbeng2024} for a pedagogical review. In this section, we outline this mapping and demonstrate how the resulting simplified thermodynamic description allows for solving both thermal and prethermal cycles in both the infinitesimal and finite cases. In particular, we derive the full analytical solution for the infinitesimal cycle, providing a concrete example of the general formulas in Eqs. (5) and (6).

\subsection{Jordan-Wigner mapping to free fermions} 
The Hamiltonian of the transverse one-dimensional Ising model (7) is given by 
\be
H_\text{Ising}=-\sum_{j} \left(\sigma^x_{j+1}\sigma_j^x+h\sigma_j^z\right)\, ,
\ee
where $\sigma^{x,y,z}$ are the standard Pauli matrices. The single-site Hilbert space is two-dimensional, and the two states are connected by the $SU(2)$ ladder operators $\sigma^\pm=(\sigma^x\pm i\sigma^y)/2$.
One can introduce fermionic creation/annihilation operators, ${c}_i^\dagger,{c}_i$, via the Jordan-Wigner transformation \cite{Mbeng2024} $
\exp\{i\pi\sum_{i<j}{c_i}^\dagger{c_i} \}{c_j}^\dagger = \sigma_j^+$, 
which satisfy the canonical anticommutation relations, $\{c_j^\dagger,c_i\}=\delta_{i,j}$ and $\{c_j^\dagger,c_i^\dagger\}=\{c_j,c_i\}=0$. This transformation maps the Hamiltonian into
\be
H = -\sum_{j} \left(c_j^\dagger c_{j+1} + c_j^\dagger c_{j+1}^\dagger + \text{h.c.} \right) + h\sum_{j=-L}^{L} \left(2c_j^\dagger c_j -1 \right).
\ee
This Hamiltonian is quadratic and can be diagonalized in the Fourier basis through a Bogoliubov rotation:
\be
\begin{pmatrix}
c_j \\  
c_j^\dagger \\
\end{pmatrix}=
\int_{-\pi}^\pi\frac{\dd\lambda}{\sqrt{2\pi}}e^{i\lambda j}U_{\theta_\lambda}
\begin{pmatrix}
\gamma(\lambda) \\  
\gamma^\dagger(-\lambda) \\
\end{pmatrix},
\qquad
U_{\theta_\lambda} = 
    \begin{pmatrix} 
        \cos \theta_\lambda & i \sin \theta_\lambda \\ 
        i \sin \theta_\lambda & \cos \theta_\lambda
    \end{pmatrix}
\ee
where ${\gamma}(\lambda)$ are canonical fermionic operators $\{ {\gamma}(\lambda), {\gamma}^\dagger(\mu) \} = \delta(\lambda - \mu)$. The angle $\theta_\lambda$ parameterizes the Bogoliubov rotation, and the choice $    \theta_\lambda = -\tfrac{1}{2i} \log \left( \tfrac{h - e^{i\lambda}}{(\cos \lambda - h)^2 + \sin^2 \lambda} \right)$
diagonalizes the Hamiltonian as $H(h) = \int \dd\lambda \, e(\lambda,h) {\gamma}^\dagger(\lambda) {\gamma}(\lambda)+\text{const.}$ with $e(\lambda,h) = 2 \sqrt{(\cos \lambda - h)^2 + \sin^2 \lambda}$ within the Brillouin zone $[-\pi,\pi]$. In contrast with the main text, we make the dependence on $h$ explicity for the sake of clarity. From the new fermionic ladder operators, an infinite number of conserved quantities can be constructed $n(\lambda)={\gamma}^\dagger(\lambda) {\gamma}(\lambda)$ \cite{Mbeng2024}.
\\

\subsection{GGEs and adiabatic flow equations} 
A thermal state at inverse temperature $\beta$ is described by the Fermi-Dirac momentum density $\rho(\lambda,h)=\frac{1}{2\pi}\left( 1+e^{\beta e(\lambda,h)}\right)^{-1}$. It can be shown that any momentum density $\rho(\lambda,h)$ uniquely corresponds to a generalized Gibbs ensemble (GGE) \cite{Mbeng2024}. In particular, if $\rho(\lambda,h)$ cannot be expressed in the Fermi-Dirac functional form, it corresponds to a non-thermal GGE.
According to Eq. (2), the evolution of $\beta$ along the thermal adiabatic stroke is given by $\langle H^2\rangle_\text{c} \partial_h\beta+\langle H\partial_hH\rangle_\text{c}\beta=0$, where the correlators are computed as in Eq. (12). The prethermal adiabatic flow is instead governed by the unitary evolution of the system. In fact, because of the diagonalization of the Hamiltonian, the chain behaves as a system of decoupled harmonic oscillators to which the adiabatic theorem applies. Consequently, the momentum density remains invariant along the prethermal flow. 
In particular, the application of the generalized hydrodynamics (GHD) equation (9) to the Ising chain also gives stationarity since the effective force vanishes, in agreement with the adiabatic theorem.
\\

\subsection{Infinitesimal cycles}

In order to compare the cycles, we compute the energy difference between thermal and prethermal strokes starting from a thermal state at external magnetic field $h_0$ and inverse temperature $\beta(h_0)$. The internal energy during the stroke along the prethermal states is given by $\langle H(h)\rangle^\pr=\int_{-\pi}^\pi \dd\lambda\, e(\lambda,h)\frac{1}{2\pi}\left( 1+e^{\beta(h_0)e(\lambda,h_0)}\right)^{-1}$, whereas for the thermal case, $\langle H(h)\rangle^\thr=\int_{-\pi}^\pi \dd\lambda\, e(\lambda,h)\rho(\lambda,h)$
where $\rho(\lambda,h)=\frac{1}{2\pi}\left( 1+e^{\beta(h)e(\lambda,h)}\right)^{-1}$ and $\beta(h)$ is the solution of the thermal adiabatic flow equation. 
When expanding in small strokes of size $\delta h$, the difference at first order is zero. We obtain the second-order correction
\be
\langle H(h_0+\delta h)\rangle^\pr - \langle H(h_0+\delta h)\rangle^\thr =
-\frac{\delta h^2}{2}\int_{-\pi}^\pi \frac{\dd\lambda}{2\pi} \left( 2\partial_he(\lambda,h_0)\partial_h\rho(\lambda,h_0) + e(\lambda,h_0)\partial_h^2\rho(\lambda,h_0) \right).
\ee
One then integrates by parts the second term and uses
explicitly the flow equation to evaluate the first derivative of the momentum density. Recollecting everything in terms of correlators (see Sec. \ref{sec_ad_flow}).
\be
\langle H(h_0+\delta h)\rangle^\pr - \langle H(h_0+\delta h)\rangle^\thr =
\beta\frac{\delta h^2}{2}\langle \partial_h H|\mathbb{P}^\pr|\partial_h H\rangle
\left( 1 - \frac{|\langle H\partial_hH\rangle_\text{c}|^2}{\langle H^2\rangle_\text{c}\langle \partial_h H|\mathbb{P}^\pr|\partial_h H\rangle}  \right),
\ee
where the correlators are clearly computed in the initial thermal state and $\langle \partial_h H|\mathbb{P}^\pr|\partial_h H\rangle=\int_{-\pi}^\pi \dd\lambda\,(\partial_he(\lambda,h_0))^2\frac{1}{2\pi}\left( 1+e^{\beta(h_0)e(\lambda,h_0)}\right)^{-1}$ (see Eq. \eqref{eq_dchprth}).

To compare this formula with Eq. (6), which expresses the difference between the extracted works, we must also consider the stroke that starts from the other reservoir in the infinitesimal cycle, at \((h_0+\delta h,\beta(h_0)+\delta\beta)\). The expression for the energy difference at the end of this stroke remains the same, except that the correlators must be evaluated at \((h_0+\delta h,\beta(h_0)+\delta\beta)\), and the stroke direction is reversed, \(\delta h \to -\delta h\). The first modification introduces only subleading corrections, while the second leaves the formula unchanged. Taking everything into account, the work difference \(\delta\mathcal{W}=\mathcal{W}^\pr-\mathcal{W}^\thr\) is simply minus twice the previous expression, in full agreement with Eq. (6).  In this case, where the thermal space is one-dimensional, it appears immediately that the sign of the above formula depends exclusively on the temperature. The autocorrelator and $\langle \partial_h H|\mathbb{P}^\pr|\partial_h H\rangle$ are clearly positive, and the term in round brackets can be bounded using a simple Cauchy-Schwarz inequality.
With similar computations, one can derive the leading contributions to the extracted work either from the thermal or prethermal cycle
\be
\mathcal{W}^\thr = -\delta h\delta\beta\langle H\partial_hH\rangle_\text{c}-\beta\delta h^2\frac{|\langle H\partial_hH\rangle_\text{c}|^2}{\langle H^2\rangle_\text{c}}\, ,\hspace{2pc}
\mathcal{W}^\pr = -\delta h\delta\beta\langle H\partial_hH\rangle_\text{c}-\beta\delta h^2\langle \partial_h H|\mathbb{P}^\pr|\partial_h H\rangle\, .
\ee
Notice that the second term of the thermal work can be written using the thermal projector, which is one-dimensional in this case: $\mathbb{P}^\thr=[\langle H^2\rangle_\text{c}]^{-1}|H\rangle\langle H|$. Specifically, it takes the form $\langle \partial_h H|\mathbb{P}^\thr|\partial_h H\rangle$ (see Methods in the main text).
In general, the work depends on the shape of the cycle and cannot be strictly bounded to be larger than zero, as required for a heat engine. A simplification is introduced if we consider skewed cycles where \( |\delta\beta|\gg|\delta h| \). Consequently, the difference between the two cycles becomes subleading; the prethermal work coincides with the thermal one $\mathcal{W}^\pr\simeq\mathcal{W}^\thr\simeq-\delta h\delta\beta\langle H\partial_hH\rangle_\text{c}$,
and this can always be made positive by tuning the sign of the variations. Furthermore, the heat absorbed in a skewed cycle is given by $\mathcal{Q}_\text{abs}\simeq|\delta\beta|\langle H^2\rangle_\text{c}$, and at leading order does not distinguish between the two cycles.
The heat engine efficiency in the skewed thermal cycle takes a particularly simple form
\be
\eta^\thr\simeq\frac{|\delta h\langle H\partial_hH\rangle_\text{c}|}{\langle H^2\rangle_\text{c}}\, ,
\ee
and the prethermal efficiency differs from the thermal one according to
\be
\frac{\eta^\pr}{\eta^\thr} \simeq 1 - \biggl|\frac{\delta h}{\delta\beta}\biggr|\beta\frac{\langle\partial_hH|\mathbb{P}^\pr|\partial_hH\rangle}{|\langle H\partial_hH\rangle_\text{c}|}\left( 1 - \frac{|\langle H\partial_hH\rangle_\text{c}|^2}{\langle H^2\rangle_\text{c}\langle\partial_hH|\mathbb{P}^\pr|\partial_hH\rangle}  \right)\, .
\ee
The last two formulas are plotted in panels a) and b) of Fig. 2 in the main text. In these plots, only negative temperature is reported, since due to particle-hole symmetry of free fermionic theories on the lattice, at positive temperature, the plot would be the same, with a reversed sign in the one for the relative efficiency.

\subsection{Finite cycles}
\begin{figure*}[t!]
\includegraphics[width=0.99\textwidth]{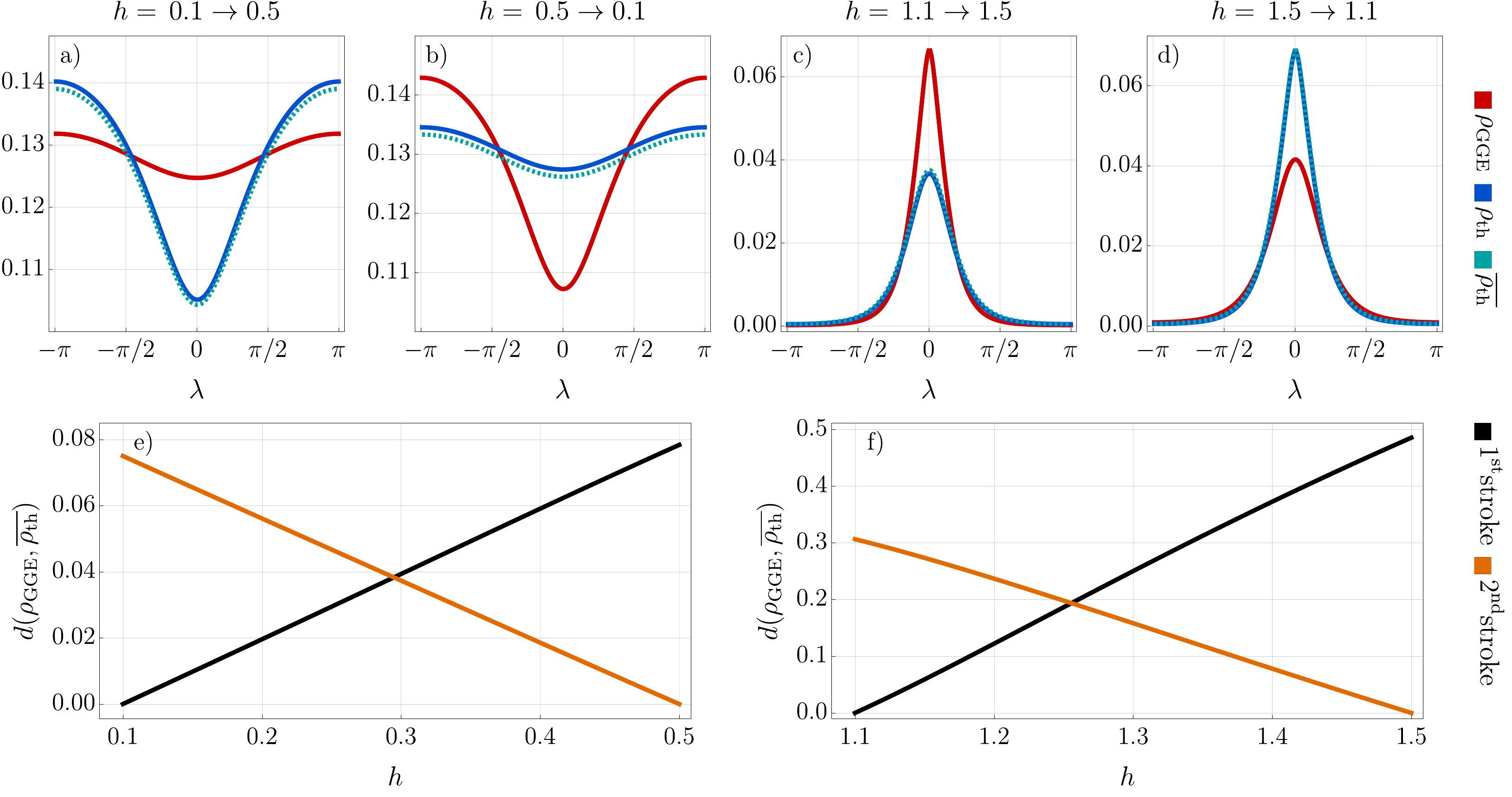}
\caption{\textbf{Comparison between the momentum distributions in the Ising model at the end of finite strokes and relative distance of GGEs from equlibrium. } Panel (a-d): Ising model's momentum distributions $\rho(\lambda)$ at the end of the strokes in the cycles in Fig. 3 of the main text. In each panel, three momentum distributions are plotted: $\rho_\text{GGE}$ is the one unitary evolved, equal to the initial thermal state due to the adiabatic theorem. The final state along the thermal stroke is $\rho_\text{th}$, while $\overline{\rho_\text{th}}$ is thermal with the same energy as the prethermal one. The latter is useful both for quantifying non-ergodicity and comparing the two cycles.
Panels a) and b) refer to strokes in the negative temperature cycle, as one can see since the more energetic modes are excited, while panels c) and d) refer to the positive temperature cycle. In this case, despite the thermal momentum distributions are not very different, an appreciable energy difference appears (see Fig. 3. Panel (e,f): evolution along the stroke of the GGE's distance from equilibrium, measured as the relative distance in the $L_2$ norm between $\rho_\text{GGE}$ and $\overline{\rho}_\text{th}$. For the explicit formula, see Eq. \eqref{eq_GGEdist}. The orange and the black lines refer to the two different strokes of the prethermal cycle. The maximum relative distance goes from the $8\%$ of the strokes at negative temperature, up to $50\%$ in the first stroke at positive temperature, meaning the system is considerably driven out of equilibrium.}
\label{Fig_Isingfillings}
\end{figure*}
Finite cycles are numerically solved according to Section \ref{sec_num}. In the case of the Ising chain, the solution of the prethermal cycle is particularly simple since the momentum density does not evolve. The results are reported in Fig. 3 of the main text.
The results for infinitesimal cycles guided us in identifying the regions of the parameter space where the difference between the two strokes was greater. In particular, we find as the most informative quantity $\delta\mathcal{W}/(\delta h\langle\partial_hH\rangle)$, which tells how much the thermal and prethermal strokes differ from each other, relative to the energy change. Based on this reasoning, we identify the cycles reported in the main text. 

As a complement to those plots, Fig. \ref{Fig_Isingfillings} presents a comparison of the system's momentum distributions at the end of each stroke. For the two strokes at negative temperature, the difference is more pronounced, as expected from the energy difference shown in Fig. 3.
For the strokes at positive temperature, we selected a set of parameters that produces a milder difference. A larger discrepancy would have resulted in negative extracted work in the prethermal cycle, meaning that it would no longer work as a heat engine. Even with these small differences, the efficiency of the thermal cycle remains an order of magnitude higher. Furthermore, we present in Fig. \ref{Fig_Isingfillings} the evolution along the prethermal strokes of the GGE's distance from equilibrium, measured as the relative distance in the $L_2$ norm between $\rho(\lambda)$ describing the GGE and $\bar{\rho}_{\thr}(\lambda)$ of a thermal state with the same energy than the GGE (i.e. the thermal state one would relax to from the GGE without further work injection). Explicitly, the formula takes the form
\be\label{eq_GGEdist}
d(\rho_\text{GGE},\bar{\rho}_\text{th})=\left(\frac{\int \dd\lambda [\rho_{\text{GGE}}(\lambda)-\bar{\rho}_{\text{th}}(\lambda)]^2}{ \int \dd\lambda [\bar{\rho}_{\text{th}}(\lambda)]^2}\right)^{1/2}\, .
\ee
From these plots we can read that the system is considerably driven out of equilibrium: at the end of the strokes at negative temperature the relative distance is around the $8\%$, and for the first stroke at positive temperature goes up to $50\%$. These plots give a quantitative evidence that we are considering a genuine nonequilbrium phenomenon.

We conclude this section by noting that although the bound on infinitesimal cycles does not constitute formal proof, we have tested a large number of finite cycles and consistently found that prethermal cycles are more efficient at negative temperatures and vice versa at positive temperatures.

\section{The XXZ spin chain in easy-axis regime}
\label{sec_XXZ}

In this section, we provide an overview of the thermodynamics and GHD of the XXZ spin chain, discuss the equations for infinitesimal cycles, and further elaborate on finite cycles.

\subsection{Thermodynamic Bethe ansatz}
We refer mainly to Ref. \cite{takahashi2005thermodynamics}; see also Ref. \cite{Bastianello2019} for conventions.
The Hamiltonian is given by
\be
H_\text{XXZ}=-J\sum_j\left(\sigma^x_{j+1}\sigma_j^x+\sigma_{j+1}^y\sigma_j^y+\Delta \sigma^z_{j+1}\sigma^z_j\right)\, ,
\ee
where $\sigma^{x,y,z}$ are the standard Pauli matrices. For anisotropy values $|\Delta|>1$, spin alignment is favored along the $z$-axis in the so-called easy-axis regime. The sign of $J\Delta$ distinguishes between the ferromagnetic ground state (if positive) and the N\'eel state (if negative). 
The many-body Hilbert space can be described in terms of excitations over the ferromagnetic ground state. The fundamental excitations are magnons --isolated and delocalized spin flips-- and bound states thereof, which within the TBA framework are referred to as ``Bethe strings" or ``strings", in short. From the perspective of the Thermodynamic Bethe Ansatz (TBA) \cite{takahashi2005thermodynamics}, each string is treated as a distinct quasi-particle species, each associated with its own root density $\{\rho_j(\lambda)\}_{j=1}^\infty$, filling function $\{\vartheta_j(\lambda)\}_{j=1}^\infty$, and pseudoenergy $\{\varepsilon_j(\lambda)\}_{j=1}^\infty$. The system is thus described by coupled TBA equations, where the scattering phase is generalized to a matrix $\Theta_{j,k}(\lambda)$ that accounts for the scattering between different types of strings.

We focus on the case with $J=-1$ and $\Delta>1$, as the Hamiltonian is symmetric under $(J,\Delta)\to(-J,-\Delta)$, meaning that for other parameter choices, the eigenproblem remains the same except for a possible sign change in energy eigenvalues.
The expectation value of the conserved charges is computed by summing the contributions of each string. As an example, the energy and momentum densities for each string species are given by
\be
e_j(\lambda)=-\frac{1}{2} \sinh(\theta) \partial_\lambda p_j(\lambda)\, , \hspace{2pc} p_j(\lambda)=2 \arctan\left[\coth\left(\frac{j \theta}{2}\right)\tan \lambda\right]\, ,
\ee
where the rapidities are constrained within the Brillouin zone $\lambda\in [-\pi/2,\pi/2]$, and the angle $\theta$ parameterizes the anisotropy as $\Delta=\cosh\theta$. The expectation values of the Hamiltonian and local magnetization are then expressed as
\be
\frac{1}{L}\langle H\rangle=\frac{J\Delta}{4}+\sum_j \int_{-\pi/2}^{\pi/2} \dd\lambda\, e_j(\lambda) \rho_j(\lambda)\, , \hspace{2pc}\langle \sigma_i^z\rangle=1-\sum_j\int_{-\pi/2}^{\pi/2} \dd\lambda\, j\rho_j(\lambda)\, ,
\ee
where $L$ is the length of the chain and $N=L - \sum_i \sigma_i^z$ is the magnons' number operator, which commutes with the Hamiltonian. Above, $J\Delta/4$ is the energy of the fully polarized state: since a state-independent offset does not affect thermodynamics, neither work nor exchanged heat, we herein neglect this additional term.
In general, higher-order correlators require summing the contributions from all string species.
The explicit form of the scattering phase is
\be
\Theta_{j,k}(\lambda)=(1-\delta_{j,k})\frac{p_{|j-k|}(\lambda)}{2\pi}+\frac{p_{j+k}(\lambda)}{2\pi}+2\sum_{\ell=1}^{\min(j,k)-1}\frac{p_{|j-k|+2\ell}(\lambda)}{2\pi}\, .
\ee

\subsection{Adiabatic flow equations}
\label{sec_ad_flow}

In this subsection, we derive the expressions for the correlators required to compute the exact flow equations (2). To simplify the notation, we consider formulas for a generic integrable system without Bethe strings: the final results are easily generalized to the XXZ model by adding a summation over the quasiparticle species.
Starting from the expectation value of a conserved charge $\tfrac{1}{L}\langle Q_j\rangle = \int\dd\lambda q_j(\lambda)\rho(\lambda)$, this can be rewritten as
\be
\frac{1}{L}\langle Q_j\rangle=\int \dd\lambda\dd\lambda' q_j(\lambda) [\vartheta^{-1}+\tfrac{1}{2\pi}\varphi]^{-1}_{(\lambda,\lambda')}\frac{1}{2\pi}\partial_{\lambda'} p
\ee
since $\rho(\lambda)=\vartheta(\lambda)\tfrac{(\partial_\lambda p)^\text{dr}}{2\pi}$ and $[1+\tfrac{1}{2\pi}\varphi\vartheta]^{-1}_{(\lambda,\lambda')}$ is the dressing integral kernel operator. The connected two-point correlator of two charges is obtained as $\langle Q_iQ_j\rangle_\text{c}=-\partial_{\beta_i}\langle Q_j\rangle$. The derivation acts only on the integral kernel as 
\be
\frac{1}{L}\partial_{\beta_i}\langle Q_j\rangle=\int \dd\lambda\dd\lambda'\dd\lambda'' q_j(\lambda) [\vartheta^{-1}+\tfrac{1}{2\pi}\varphi]^{-1}_{(\lambda,\lambda'')} \frac{\partial_{\beta_i} \vartheta(\lambda'')}{\vartheta^2(\lambda'')}[\vartheta^{-1}+\tfrac{1}{2\pi}\varphi]^{-1}_{(\lambda'',\lambda')}\frac{1}{2\pi}\partial_{\lambda'}p\, .
\ee
The derivative of the filling is $\partial_{\beta_i}\vartheta=-\vartheta(1-\vartheta)q_i^\text{dr}$, which can be verified by differentiating the TBA equation. The two integral kernels, combined with $\vartheta^{-2}(\lambda'')$, act as dressing operations, respectively, on the left and on the right. Summing everything, we obtain
\be\label{eq_chargecorrSM}
\tfrac{1}{L}\langle Q_iQ_j\rangle_\text{c}\overset{L\to\infty}{\simeq}\int \dd\lambda \, q_i\dr \rho(1-\vartheta)q_j\dr,
\ee
where $\simeq$ denotes equality in the thermodynamic limit. The derivation of the susceptibility matrix follows similar steps. The expectation value of the derivative of a charge is $\tfrac{1}{L}\langle \partial_\pp Q_j\rangle\simeq\int \dd\lambda \left(\partial_\pp q_j \rho +\frac{1}{2\pi}\partial_\lambda q_jf\dr\vartheta\right)$ \cite{Bastianello2019} , and similarly to before, can be rewritten as
\be
\tfrac{1}{L}\langle \partial_\pp Q_j\rangle\simeq\int \dd\lambda\dd\lambda'\frac{1}{2\pi} \left(\partial_\pp q_j[\vartheta^{-1}+\tfrac{1}{2\pi}\varphi]^{-1}_{(\lambda,\lambda'')}\partial_{\lambda'}p +\partial_\lambda q_j[\vartheta^{-1}+\tfrac{1}{2\pi}\varphi]^{-1}_{(\lambda,\lambda'')}f\right)\, .
\ee
The derivative follows the same steps as before but introduces an extra term because $f(\lambda)= -\partial_\pp p(\lambda) + \int \dd\lambda'\partial_\pp\Theta(\lambda-\lambda')\vartheta(\lambda')(\partial_{\lambda'} p)\dr(\lambda')$ depends on the Lagrange multiplier via the filling inside the integral
\be
-\tfrac{1}{L}\partial_{\beta_i}\langle \partial_\pp Q_j\rangle\simeq\int \dd\lambda\rho(1-\vartheta) q_i\dr\left((\partial_\pp q_j)\dr +\frac{(\partial_\lambda q_j)\dr}{(\partial_\lambda p)\dr}f\dr\right) + \int \dd\lambda (\partial_\lambda q_j)\dr\partial_{\beta_i}f\vartheta\, .
\ee
Explicitly computing $\partial_{\beta_i}f(\lambda)$,  the second integral can be absorbed into the first using $\Lambda_j(\lambda)= -\partial_\pp q_j(\lambda) + \int \dd\lambda'\partial_\pp\Theta(\lambda-\lambda')\vartheta(\lambda')(\partial_{\lambda'} q_j)\dr(\lambda')$, leading to
\be\label{eq_S18}
\tfrac{1}{L}\langle Q_i\partial_\pp Q_j\rangle_\text{c}\simeq\int \dd\lambda \, q_i\dr \rho(1-\vartheta)\left(f\dr\frac{(\partial_\lambda q_j)\dr}{(\partial_\lambda p)\dr}-\Lambda_j\dr\right)\, .
\ee
Although we can compute the flow equations exactly, evaluating the prethermal work is cumbersome due to the infinite-dimensional space of conserved charges in integrable systems. However, we can use hydrodynamic projections and consider the more accessible formula $\left[A_\pr^T C_\pr^{-1}A_\pr\right]_{i,j}=\langle \partial_\pp Q_i|\mathbb{P}^\pr|\partial_\pp Q_j\rangle$, where the scalar product is defined in the space of conserved charges as $\langle Q_i|Q_j\rangle\equiv\langle Q_i Q_j\rangle_\text{c}=L[C_\pr]_{i,j}$, and the projector is $\mathbb{P}^\pr=\sum_{i,j}[C_\pr^{-1}]_{i,j}|Q_i\rangle\langle Q_j|$. It is convenient to move to a rapidity basis $|\lambda\rangle$ with the definition $\langle Q_i|\lambda\rangle=Lq_i(\lambda)$, then in this basis the covariance matrix is written as
$\langle \lambda'|C_\pr|\lambda\rangle=\int \dd\lambda'' [1+\tfrac{1}{2\pi}\vartheta\varphi]^{-1}_{(\lambda',\lambda'')}\rho(\lambda'')(1-\vartheta(\lambda''))[1+\tfrac{1}{2\pi}\varphi \vartheta]^{-1}_{(\lambda'',\lambda)}$. In this basis, it is easy to invert $C$ and by considering its action on Eq. \eqref{eq_S18}, it follows that
\be\label{eq_dchprth}
\langle \partial_\pp Q_i|\mathbb{P}^\pr|\partial_\pp Q_j\rangle=\int \dd\lambda\, \left(f\dr\frac{(\partial_\lambda q_i)\dr}{(\partial_\lambda p)\dr}-\Lambda_i\dr\right)\rho(1-\vartheta)\left(f\dr\frac{(\partial_\lambda q_j)\dr}{(\partial_\lambda p)\dr}-\Lambda_j\dr\right)\, .
\ee

\subsection{Infinitesimal cycles}
\begin{figure*}[t!]
\includegraphics[width=0.7\textwidth]{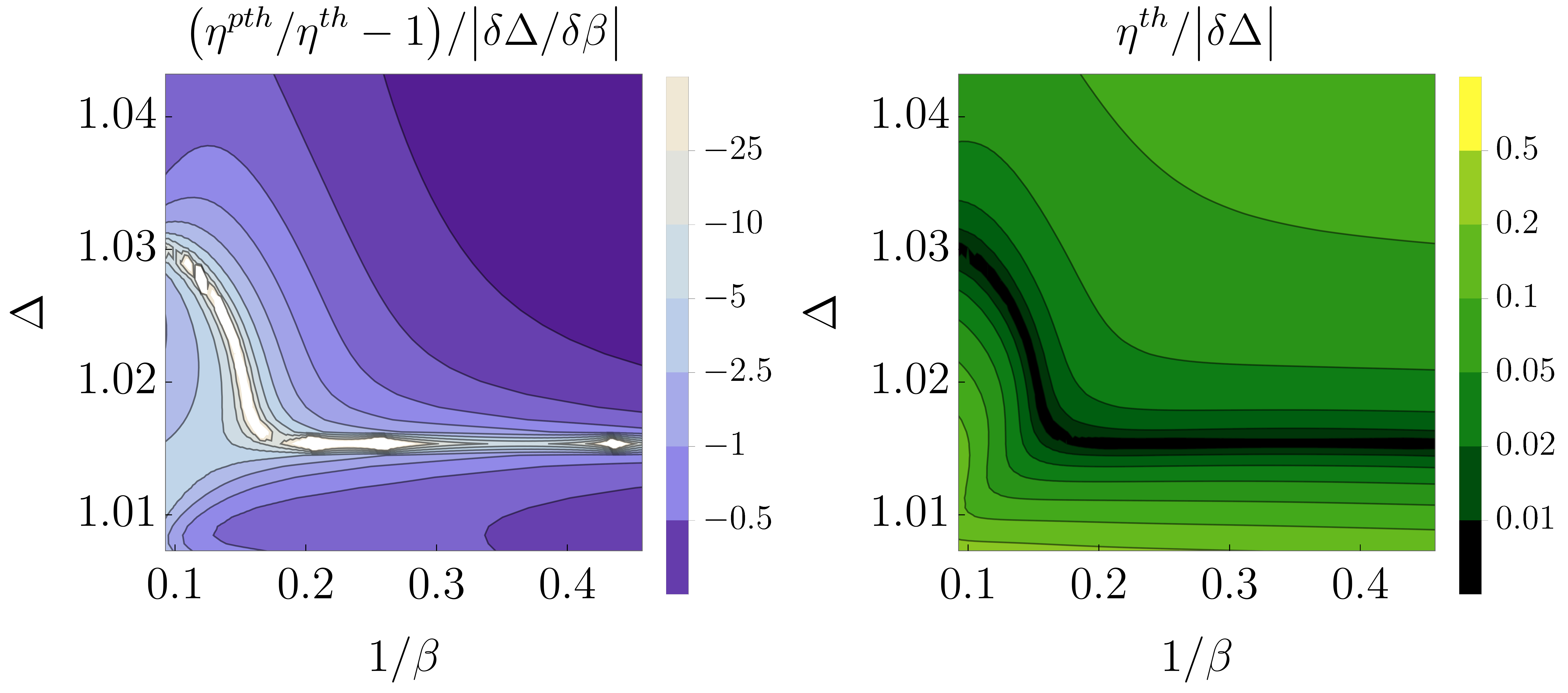}
\caption{\textbf{Thermal vs Prethermal infinitesimal Otto cycle in the XXZ chain. } Relative efficiency of thermal-prethermal infinitesimal skewed Otto cycles for the easy-axis antiferromagnetic $J=-1$ XXZ chain from Eq. \eqref{eq_releffXXZ}. The plot is at a small fixed average magnetization $\langle\sigma^z\rangle=0.05$ to enhance the string formation at small positive temperatures. As for negative temperatures, regions of large enhancement correspond to vanishing thermal efficiency.}
\label{Fig_infXXZ}
\end{figure*}

Compared to the Ising case, the equations for the XXZ chain should also take into account the conservation of total magnetization, even for thermal evolution. Before stating the results for this specific case, let us go back to the general case discussed in the main text, where the prethermal cycle conserves $N$ charges, the thermal one $N'<N$, and the tunable parameter is $\chi$. From Eq. (5) follows that the extracted works are
\be
\mathcal{W}^{\thr/\pr} = \delta\pp\delta\beta_i\partial_{\beta_i}\langle\partial_\pp H\rangle-\beta (\delta\pp)^2 \left[A^T_{\thr/\pr} C^{-1}_{\thr/\pr}A_{\thr/\pr}\right]_{H,H} \, ,
\ee
where the notation is the same as in Eq. (6) and we sum over repeated indices up to $N'$, since the system can only exchange the thermal charges with an external reservoir. In the assumption that reservoirs exchange only energy with the system and not other conserved charges, the result is further simplified. If so, the $\delta\beta_i$ are not independent, as we must impose to conserve the thermal charges between the two strokes. Hence, $\delta\beta_i\partial_{\beta_i}\langle Q_j\rangle=0$ with $j=2,\ldots,N'$, where the energy $Q_1\equiv H$ is excluded.
The XXZ chain conserves only the number of magnons (i.e., the magnetization) in addition to the energy, obtaining
\be
\mathcal{W}^{\thr/\pr} = -\delta\pp\delta\beta \left(\langle H\partial_\Delta H\rangle_\text{c} - \frac{\langle N\partial_\Delta H\rangle_\text{c}\langle NH\rangle_\text{c}}{\langle N^2\rangle_\text{c}} \right)-\beta (\delta\pp)^2 \left[A^T_{\thr/\pr} C^{-1}_{\thr/\pr}A_{\thr/\pr}\right]_{H,H}\, ,
\ee
where we identify $\delta\beta\equiv\delta\beta_1$. Notice that this result can be consistently obtained by a brute-force expansion of the GHD and TBA equations, as it should. However, this route is longer than the derivation shown above, and therefore, we do not report it. As discussed in the Methods section of the main text, the second term of the work can be expressed in terms of projectors onto the space of conserved charges, i.e., $\left[A^T_{\thr/\pr} C^{-1}_{\thr/\pr}A_{\thr/\pr}\right]_{H,H}=\langle\partial_\Delta H|\mathbb{P}^{\thr/\pr}|\partial_\Delta H\rangle$.
As we did for the Ising model in Section \ref{sec_ising}, we focus on skewed cycles to achieve further simplifications.
The leading absorbed heat does not distinguish the two cycles and is given by $\mathcal{Q}_\text{abs}\simeq|\delta\beta_i\partial_{\beta_i}\langle H\rangle|$, and the infinitesimal efficiencies become
\be
\eta^\thr \simeq \frac{\biggl|\delta\Delta \left(\langle H\partial_\Delta H\rangle_\text{c} - \frac{\langle N\partial_\Delta H\rangle_\text{c}\langle NH\rangle_\text{c}}{\langle N^2\rangle_\text{c}} \right) \biggr|}{\langle HH\rangle_\text{c} - \frac{|\langle NH\rangle_\text{c}|^2}{\langle N^2\rangle_\text{c}} }\, ,
\ee
and
\be\label{eq_releffXXZ}
\frac{\eta^\pr}{\eta^\thr} \simeq 1 - \biggl|\frac{\delta \Delta}{\delta\beta}\biggr|\beta\frac{\langle\partial_\Delta H|\mathbb{P}^\pr|\partial_\Delta H\rangle}{\biggl|\langle H\partial_\Delta H\rangle_\text{c} - \frac{\langle N\partial_\Delta H\rangle_\text{c}\langle NH\rangle_\text{c}}{\langle N^2\rangle_\text{c}}\biggr|}\left( 1 - \frac{|\langle H\partial_\Delta H\rangle_\text{c}|^2}{\langle H^2\rangle_\text{c}\langle\partial_\Delta H|\mathbb{P}^\pr|\partial_\Delta H\rangle} - \frac{|\langle N\partial_\Delta H\rangle_\text{c}|^2}{\langle N^2\rangle_\text{c}\langle\partial_\Delta H|\mathbb{P}^\pr|\partial_\Delta H\rangle} \right)\, ,
\ee
and again it is easy to see that the prethermal is more efficient at negative temperature due to the Bessel inequality. All terms appearing in the above formula are explicitly reported in Section \ref{sec_ad_flow}.

 These formulas are those plotted in Fig.2. Fig. \ref{Fig_infXXZ} shows a plot for positive temperature, illustrating that the thermal cycle is more efficient than the prethermal one. The regions where the differences are most striking correspond to those with vanishing thermal efficiency. Compared to the negative temperature case, we have decreased the average magnetization to $\langle\sigma^z\rangle = 0.05$ to increase the number of magnons and facilitate string formation. Since strings are thermally activated and the system conserves the total number of excitations for each string independently during the dynamics, the presence of more strings drives the system further from a thermal state, thereby enhancing the differences.
 
\subsection{Finite cycles}
\begin{figure*}[t!]
\includegraphics[width=0.99\textwidth]{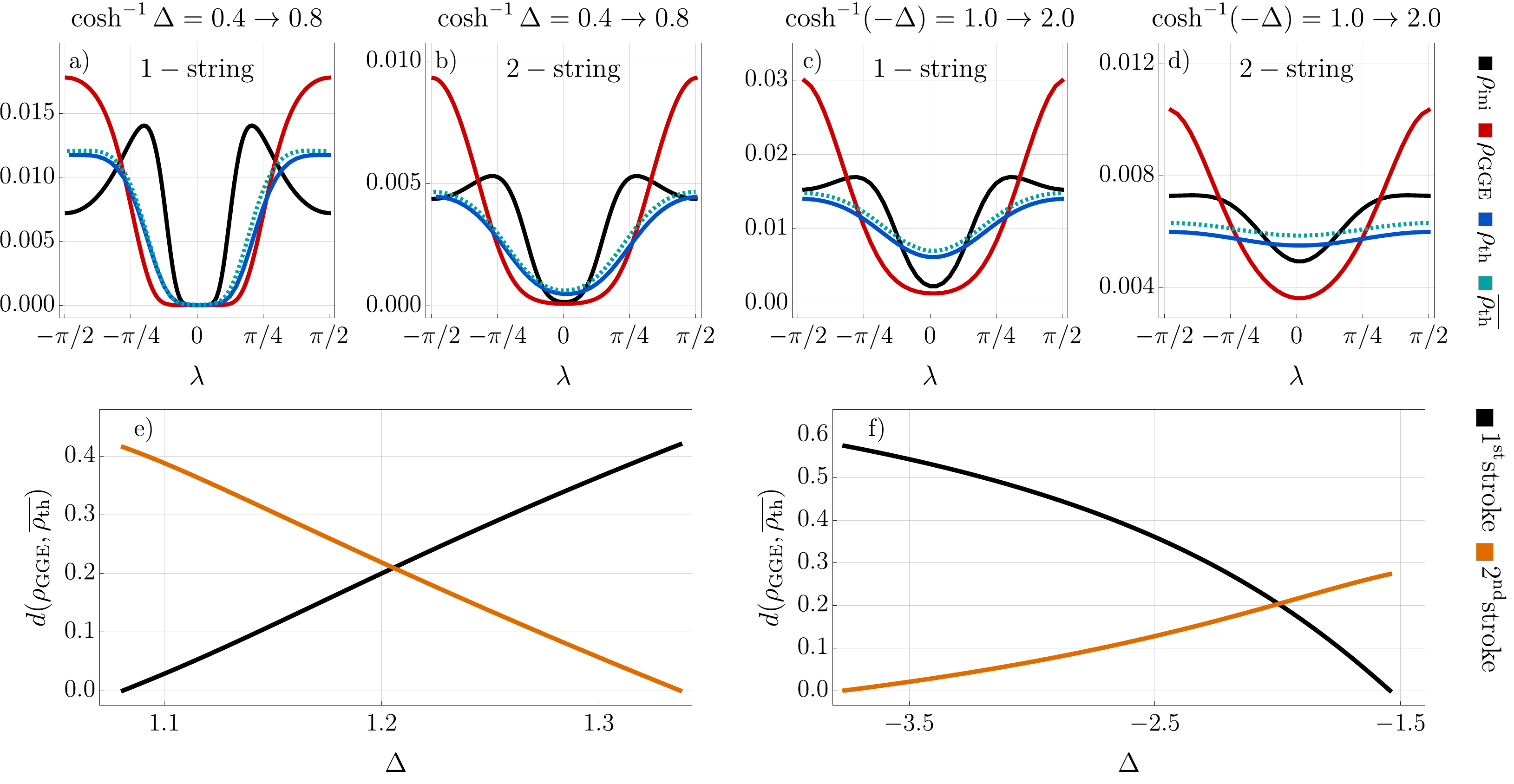}
\caption{\textbf{Comparison between root densities in the XXZ chain at the end of finite strokes and relative distance of GGEs from equlibrium. }. Panels (a) and (b) show the root densities for the first two strings, respectively, in the lowest energy stroke of the negative temperature cycle in Fig. 3. Panels (c) and (d) display the same quantities for the positive temperature cycle. In each panel, four root densities are plotted: $\rho_\text{ini}$ is the initial thermal state, $\rho_\text{GGE}$ is the one unitary evolved, and $\rho_\thr$ is the final state along the thermal stroke. The dashed lines represent $\overline{\rho_\thr}$, the root density corresponding to a thermal state with the same energy and magnetization as the one unitarily evolved via GHD. Even from the first two strings, it is evident that bound states are suppressed with their size.  Notably, higher modes remain excited even at positive temperature due to the unitary equivalence between \((J=-1,\Delta<-1)\) and \((J=1,\Delta>1)\). Panel (e,f): evolution along the stroke of the GGE's distance from equilibrium, measured as the relative distance in the $L_2$ norm between $\rho_\text{GGE}$ and $\overline{\rho}_\text{th}$. The explicit formula is reported in the text. The orange and the black lines refer to the two different strokes of the prethermal cycle. The maximum relative distance goes from the $40\%$ of the strokes at negative temperature, up to $60\%$ in the first stroke at positive temperature, meaning the system is considerably driven out of equilibrium.}
\label{Fig_XXZfillings}
\end{figure*}

In Fig. 3, for the plot at positive temperature, we have moved into the region of negative anisotropy $\Delta$.
Although infinitesimal cycles at positive $\Delta$ show a noticeable difference (see Fig. \ref{Fig_infXXZ}), we did not find finite cycles in which the prethermal-thermal difference was appreciable. We believe this is because achieving a sufficient number of strings to drive the system significantly away from a thermal state requires increasing the number of magnons, and consequently, the energy. As a result, the relative difference between prethermal and thermal cycles became too small. To circumvent this, we shifted our analysis to the region of negative $\Delta$.
Due to the symmetry under $(J,\Delta)\to(-J,-\Delta)$,  this choice is equivalent to considering the case $J=+1$ and $\Delta>1$. Under this transformation, the energy eigenvalues reverse their sign while the eigenvectors remain unchanged. From a thermodynamic perspective, this corresponds to reversing the sign of the temperature. This allowed us to activate more strings while keeping the energy low, even at a positive temperature.

As an additional complement, we present in Fig. \ref{Fig_XXZfillings} a plot of the evolution of the root densities along the strokes. The figure illustrates that during the adiabatic increase of $\Delta$, excitations shift toward larger rapidities. Moreover, even though the magnetization is fixed across all the states shown in the figure, the area under the root densities following thermal states appears to vary between them. This discrepancy arises because the figure only displays the first two strings, and the contribution to the magnetization gets redistributed across strings. In particular, increasing \( |\Delta| \) favors large strings.
In contrast, the prethermal evolution separately conserves the population of each string, and it shows a higher concentration of excitations in the small string sizes shown in the plot.
Finally, we present in Fig. \ref{Fig_XXZfillings} the evolution along the prethermal strokes of the distance from the prethermal state to a thermal state with the same energy and magnetization. The formula for the distance is the generalization of Eq. \eqref{eq_GGEdist} to the presence of strings, $d(\rho_\text{GGE},\bar{\rho}_\text{th})=\left(\frac{\sum_j\int \dd\lambda [\rho_{\text{GGE};j}(\lambda)-\bar{\rho}_{\text{th};j}(\lambda)]^2}{ \sum_j\int \dd\lambda [\bar{\rho}_{\text{th};j}(\lambda)]^2}\right)^{1/2}\, $
where we make the sum over all the strings explicit for the sake of clarity. As for the Ising model, the relative distance at the end of the strokes is around the $40\%$ for negative temperatures, and goes up to $60\%$ for the first stroke at positive temperature, meaning the system is considerably far from equlibrium.
Finally, we conclude by emphasizing that our study explores a wide range of finite cycles, consistently finding, beyond any numerical error, that prethermal cycles are more efficient at negative temperatures and, conversely, less efficient at positive temperatures.

\section{Numerical Methods}
\label{sec_num}

This section summarizes the numerical methods we employed for the TBA and GHD equations. We keep the discussion compact, as these are standard methods, and refer to the literature for further details.
A commented Mathematica notebook for the numerical solution of a cycle is provided on Zenodo \cite{Zenodo}.

\subsection{Solving the TBA} 
\label{sec_num_TBA}

The TBA and GHD equations are solved by discretizing the rapidity space on a grid. We consider $\{\lambda_i\}_{i=1}^N$ a discretization of $[-\pi/2,\pi/2]$ with $\lambda_i<\lambda_{i+1}$, with the convention that $\lambda_1=-\pi/2$ and $\lambda_N=\pi/2$. We choose a flat discretization, but other choices are equally valid.
Functions like energy, momentum, root density, and so forth are discretized with the midpoint rule $e_i(\lambda)\to e_{(i,j)}=e_i\left(\frac{\lambda_j+\lambda_{j+1}}{2}\right)$, where the index $j$ runs over the rapidity discretization and $i$ over the strings, which are truncated on a maximum cutoff $i\le N_\text{str}$. This truncation is standard, as strings are thermally activated and their length is exponentially suppressed.
The scattering kernel is discretized by integrating it exactly on a finite interval
\be
\varphi_{i,i'}(\lambda-\lambda')\to \varphi_{(i,j),(i',j')}\equiv \int_{\lambda_{j'}}^{\lambda_{j'+1}} \dd\lambda'\, \varphi_{i,i'}\left(\frac{\lambda_j+\lambda_{j+1}}{2}-\lambda'\right)\, .
\ee
The integral can be easily analytically computed from the definition. This discretization is convenient for integral equations in which the scattering kernel is convoluted with a smooth function, like the filling function, even if $\varphi$ has sharp variations.
As a last step, the couple of indexes $(i,j)$ is converted into a single index ${\rm n}=j+i(N-1)$: the integral TBA equations (see Methods) are then discretized as
\be
\varepsilon(\lambda) = \sum \beta_i q_i(\lambda) + \int\frac{\dd\lambda'}{2\pi}\varphi(\lambda-\lambda')\log(1+e^{-\varepsilon(\lambda')})\hspace{1pc} \to \hspace{1pc} \varepsilon_{\rm n} = \sum_i \beta_i q_{{\rm n},i}(\lambda) + \sum_{{\rm n}'}\frac{1}{2\pi}\varphi_{{\rm n},{\rm n'}}\log(1+e^{-\varepsilon_{{\rm n}'}})\, .
\ee
The discretized non-linear equations are then solved by standard methods, like the Newton–Raphson method. A similar discretization is also used for the dressing equations (defined in Methods), which become linear matrix equations and are solved with standard linear algebra methods.

\subsection{Solving the strokes} 

In integrable models, the flow equations describing infinitesimal strokes (2) are conveniently reformulated in the quasiparticle basis. In particular, the prethermal strokes are equivalent to GHD evolution \cite{Bastianello2019}, whereas for the thermalizing stroke, we repeatedly solve the TBA equations with energy evolving on the adiabatic flow $\partial_\chi\langle H \rangle=\langle \partial_\chi H\rangle$. Hereafter, we discuss the numerical solution of the GHD equation within the method of characteristics \cite{Bastianello2019}. We discuss it at the level of the exact GHD equations, as it is more transparent, and discretization in the rapidity space straightforwardly follows from Section \ref{sec_num_TBA}.
To illustrate the method, 
consider Eq. (9), which can be rewritten in terms of the fillings \cite{Bastianello2019} as a convective equation:  
\be
\partial_\chi\vartheta(\lambda,\chi) + F^\text{eff}(\lambda,\chi)\partial_\lambda\vartheta(\lambda,\chi) = 0\, .
\ee
The solution can be then implicitly expressed as:  
\be\label{eq_ghd_imp}
\vartheta(\chi',\lambda) = \vartheta(\chi, \lambda(\chi',\chi))\hspace{2pc}\text{where}\hspace{2pc}
\lambda(\chi',\chi) = \lambda - \int_\chi^{\chi'}\dd\xi\, F^\text{eff}(\lambda(\xi,\chi),\xi).
\ee  
Here, \( F^\text{eff}(\lambda,\chi) \) depends non-trivially on the filling, making the solution implicit, but it paves the way to a systematic discretization. 
A second-order $\dd\chi$ algorithm is achieved by discretizing the integral in Eq. \eqref{eq_ghd_imp} by the midpoint rule. 
To this end, one discretizes the $\chi-$evolution on integers and half-integer steps, i.e. $n\dd\chi$ and $(n+\tfrac{1}{2})\dd\chi$. Let us call $\vartheta_n(\lambda)\equiv\vartheta(n\dd \chi,\lambda)$ and $\vartheta'_n(\lambda)\equiv\vartheta((n+\tfrac{1}{2})\dd \chi,\lambda)$  the discretized fillings.
Then, one evolves $\vartheta_n\to \vartheta_{n+1}$ by approximating Eq. \eqref{eq_ghd_imp} as $\lambda((n+1)\dd\chi,n\dd\chi)=\lambda-\dd\chi F^\text{eff}(\lambda,(n+\tfrac{1}{2}\dd\chi)\Big|_{\vartheta'_{n}}$, where the effective force is computed with the shifted filling $\vartheta'_n$. Then, in the next step,  $\vartheta_{n+1}$ is used to evolve $\vartheta'_{n}\to\vartheta'_{n+1}$, and then the step is repeated.
Notice that, for consistency, the precision up to $\mathcal{O}(\dd\chi^2)$ is maintained if interpolations in the rapidity space are at least of second order as well. While the first filling $\vartheta_{n=0}$ is determined by the initial conditions, the input for the first shifted filling $\vartheta'_{n=0}$ must be approximated: this can be done by considering the evolution for $\chi$ from $0$ to $\dd \chi/2$ as an evolution on its own, discretized with a small step $\dd\chi'\ll \dd\chi$ and using first-order discretizations for the first auxiliary step; see Ref. \cite{Bastianello2019} for details.

\end{document}